\documentclass{emulateapj}

\newcommand\beq{\begin{equation}}
\newcommand\eeq{\end{equation}}
\newcommand\beqar{\begin{eqnarray}}
\newcommand\eeqar{\end{eqnarray}}

\newcommand{\cref}{C_{\rm ref}}

\begin{document}

\title{PROTOSTAR FORMATION IN MAGNETIC MOLECULAR CLOUDS BEYOND ION DETACHMENT: III.  A PARAMETER STUDY} 

\author{Konstantinos Tassis\altaffilmark{1,2} \& Telemachos Ch. Mouschovias\altaffilmark{1}}

\altaffiltext{1}{Departments of Physics and Astronomy,
University of Illinois at Urbana-Champaign, 1002 W. Green Street, Urbana, IL 61801}
\altaffiltext{2}{Department of Astronomy and Astrophysics and the 
Kavli Institute for Cosmological Physics, 
The University of Chicago, 5640 South Ellis Avenue
Chicago, IL 60637}

\begin{abstract}

In two previous papers we formulated and solved, for a fiducial set of free 
parameters, the problem of the formation and evolution 
of a magnetically supercritical core inside a magnetically subcritical parent cloud. 
The evolution was followed into the opaque phase that resulted in the formation of 
a hydrostatic protostellar core. In this paper
we present a parameter study to assess the sensitivity of the results (1) to the density 
at which the equation of state becomes adiabatic; (2) to the initial mass-to-flux 
ratio of the parent cloud; and (3) to ionization by radioactive decay of different 
nuclei ($^{40}$K and $^{26}$Al) at high densities ($n_{\rm n} \gtrsim 10^{12}$ ${\rm cm^{-3}}$). 
We find that (1) the results depend only slightly on the density 
at which the onset of adiabaticity occurs; (2) memory of the initial mass-to-flux 
ratio is completely lost at late times; and (3) the precise source of radioactive 
ionization alters the degree of attachment of the electrons to the field lines (at 
high densities), and the relative importance of ambipolar diffusion and Ohmic 
dissipation in reducing the magnetic flux of the protostar. The value of the 
magnetic field at the end of the runs is insensitive to the values of the free 
parameters and  in excellent agreement with meteoritic measurements of the 
protosolar nebula magnetic field. The magnetic flux 
problem of star formation is resolved for at least strongly magnetic newborn stars. 
A complete detachment of the magnetic field from the matter is unlikely. The formation 
of a ``magnetic wall'' (with an associated magnetic shock) is independent of the
assumed equation of state, although the process is enhanced and accelerated by the formation
of a central hydrostatic core.

\end{abstract}


\section{Introduction}

We previously formulated the problem of the
ambipolar-diffusion--initiated core formation and evolution in
self-gravitating, magnetically supported model clouds (Tassis \& Mouschovias 2006a, 
hereafter Paper I), and presented the solution for a typical 
model cloud (Tassis \& Mouschovias 2006b, hereafter Paper II). In this
paper, we quantify the dependence of the 
results on the values of the relevant free parameters of the problem. 

The system of MHD equations governing the evolution of the model
cloud,  which  we presented in Paper I, contains four free
parameters: the dimensionless initial central mass-to-flux ratio $\mu_{\rm d,c0}$ of 
the reference state relative to its critical value for collapse;  
the characteristic lengthscale $\tilde{l}_{\rm ref}$ of 
the initial column density distribution relative to the thermal critical lengthscale;
the ratio of the external thermal pressure and ``gravitational pressure''
in the central flux tube of the initial reference state, 
$\tilde{P}_{\rm ext}$; and the density at which the transition from an
isothermal to an adiabatic equation of state occurs, $n_{\rm opq}$.
Additional parameters involved in the chemical model are well constrained
observationally  and experimentally (see Appendix in Desch \& Mouschovias 2001 and
references therein), 
and values from these studies are used in the calculations.

Of the parameters referred to above, the value of $\tilde{P}_{\rm ext}$ has no 
significant effect on the formation and evolution of the supercritical core in 
a {\em self-gravitating} model cloud. The value of $\tilde{l}_{\rm  ref}$ is a 
measure of the strength of the initial thermal-pressure forces relative to 
gravitational forces inside the
model cloud. If $\tilde{l}_{\rm ref}>3\sqrt{2}$ then the cloud is
thermally supercritical. \cite{BM95a} showed that, as long as the
cloud is significantly thermally supercritical, the value of $\tilde{l}_{\rm ref}$
plays no role other than to determine the size and hence the 
mass of the initial cloud (which they found to be proportional to
$\tilde{l}_{\rm ref} ^2$). Hence, the effect of varying $\tilde{P}_{\rm ext}$
and $\tilde{l}_{\rm ref}$ is not examined further in this parameter 
study. The values used for these parameters in 
all the models presented in this paper are $\tilde{P}_{\rm ext} =
0.1$ and $\tilde{l}_{\rm ref}=5.5\pi = 4.07 \times  (3\sqrt{2})$.

The free parameters varied are the density at which the equation of state
becomes adiabatic, $n_{\rm opq}$, and the initial mass-to-flux ratio
of the parent cloud $\mu_{\rm d}$ in units of the critical value. We also
examine how the results change when, at high densities (when natural
radioactivity dominates cosmic rays as the primary ionization
mechanism), radioactive $^{26}$Al rather than $^{40}$K is the most
abundant decaying element. 

\begin{table}
\begin{center} 
\caption{\label{models} Model Parameters}
\vspace{3mm}
\begin{tabular}{lccc}
\hline\hline
Model & $n_{\rm opq} {\rm \, (cm^{-3})}$ & $\mu_{\rm d,c0}$ & dominant radionuclide \\
\hline \hline
$f$ & $10^{11}$ & $0.25$ & $^{40}$K \\ 
\hline
$n10$ & $10^{10}$ & $0.25$ & $^{40}$K \\
$n12$ & $10^{12}$ & $0.25$ & $^{40}$K \\
$i$  & always isothermal & $0.25$ & $^{40}$K\\
\hline
$m5$ & $10^{11}$ & $0.50$ & $^{40}$K \\
$m7$ & $10^{11}$ & $0.70$ & $^{40}$K \\
$m9$ & $10^{11}$ & $0.90$ & $^{40}$K \\
\hline
$Al$  &  $10^{11}$ & $0.90$ & $^{26}$Al \\
\hline \hline
\end{tabular}
\end{center}
\end{table}

Table \ref{models} shows the relevant parameters of the
models studied in this paper. Model $f$ is our fiducial run
that was presented in Paper II. In \S \, \ref{nop} we examine how
the results of the calculations change as the density $n_{\rm opq}$, 
at which the transition from isothermality to adiabaticity occurs,
is varied. For this purpose, we compare results from three
models: $n$10, $f$ and $n$12, in which all parameters
have the same values as in the fiducial run, except 
$n_{\rm opq}$; it is equal to $10^{10}$, $10^{11}$ and $10^{12} {\, \rm cm^{-3}}$ 
in the three models, respectively. These values cover the expected density range
in which the collapsing core will become optically thick
\citep{Gaustad63,Hayas66,DL84,OH94,MI99}.
As part of this parameter study, we also present in \S \, \ref{iso} results 
for model $i$, a ``control run'', in which
the isothermal approximation is retained throughout the
calculation. Although clearly unrealistic, this run is used 
to assess which effects are caused by the change in the
equation of state.
The effect of the initial mass-to-flux ratio on the solution
is discussed in \S \, \ref{mtfvar}. We present results for 
four models, $f$, $m$5, $m$7 and $m$9, with parameters always the same as
those of the fiducial run, except $\mu_{\rm d,c0}$,
which is equal to $0.25$, $0.5$, $0.7$ and $0.9$, respectively.  
Finally, in \S \, \ref{alvar} we examine how the (late-time) results
depend on the kind of radioactive element that dominates the ionization
process at high densities. If the mass-to-flux ratio of the 
parent cloud is close to critical, then the evolution is much more
rapid than in the fiducial model, and 
radionuclides with half-lives shorter than $^{40}$K
may contribute significantly to the
ionization at high densities.  For this reason, we compare models $m$9
and $Al$, both of which have $\mu_{\rm d,c0}=0.9$, but assume different
dominant radionuclides ($^{40}$K and   $^{26}$Al, respectively).

\section{Dependence on $n_{opq}$}\label{nop}

In this section we examine the effect of varying the density at which
the assumed equation of state changes from isothermal to adiabatic. We
compare three models: $n$10 ($n_{\rm opq}=10^{10}$ ${\rm cm^{-3}}$); 
 $f$ ($n_{\rm opq}=10^{11}$ ${\rm cm^{-3}}$); and
 $n$12 ($n_{\rm opq}=10^{12}$ ${\rm cm^{-3}}$). Model  
$f$ is the fiducial run, which was discussed in detail in Paper II. 
All three models have an initial mass-to-flux ratio
$\mu_{\rm d,c0} = 0.25$.

\begin{figure*}
\plotone{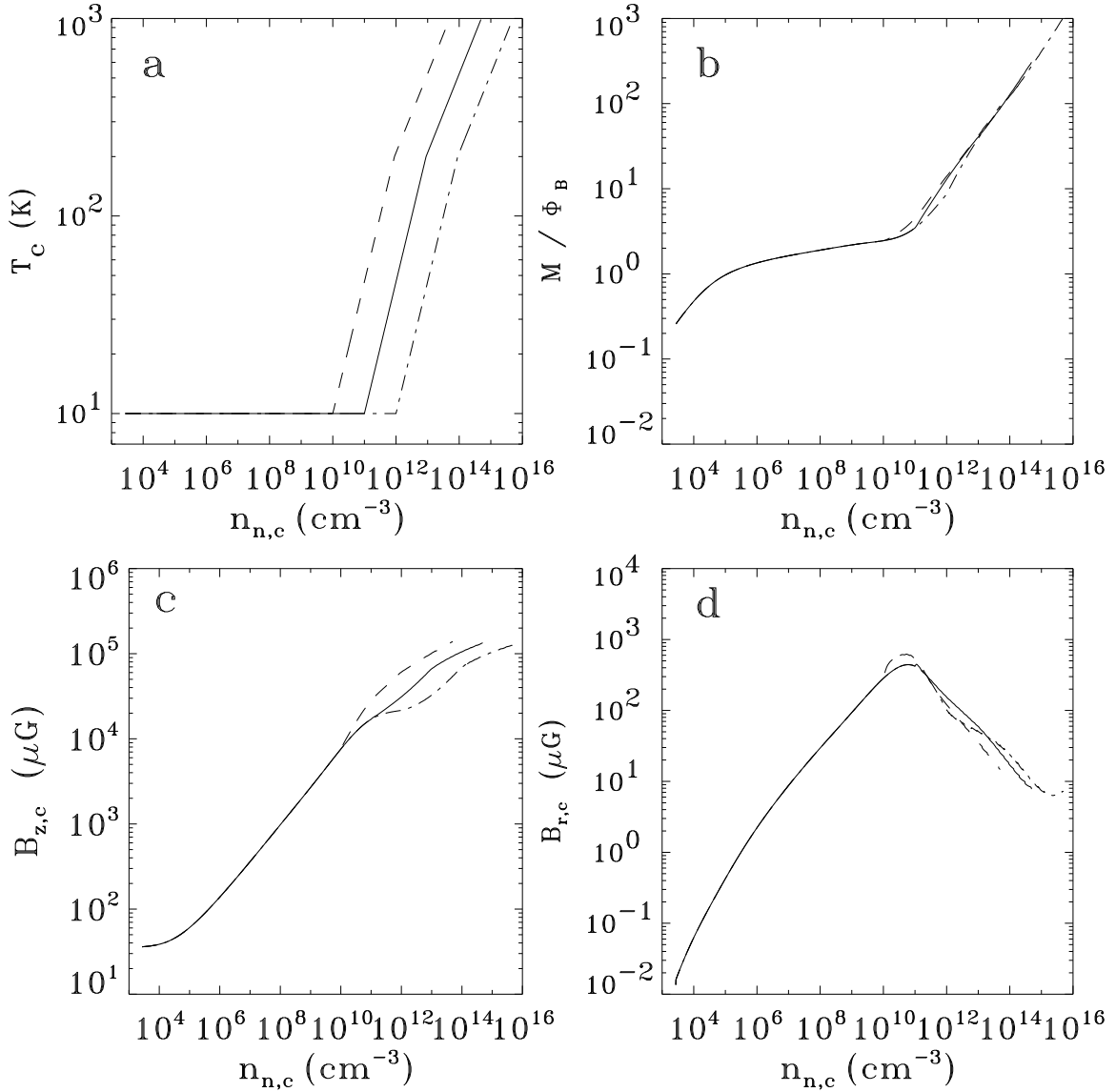}
\caption{\label{nparam_cen_a} Central quantities as functions of
the central number density of neutrals (in ${\rm cm^{-3}}$) for
models $f$ (solid line, $n_{\rm opq}=10^{11} {\, \rm cm^{-3}}$), 
$n$10 (dashed line,  $n_{\rm opq}=10^{10} {\, \rm cm^{-3}}$), 
and $n$12 (dashed-dot line, $n_{\rm opq}=10^{12} {\, \rm cm^{-3}} $). 
(a) Central temperature (in ${\rm K}$); 
(b) central mass-to-flux ratio normalized to the critical value; 
(c) $z-$component of the magnetic field (in ${\rm \mu G})$ at the center; 
(d) $r-$component of the magnetic field (in ${\rm \mu G})$ at the center.
}
\end{figure*}

Figure (\ref{nparam_cen_a}) exhibits central quantities
of the model clouds as functions of the central number density of
neutrals. In all cases, the solid line corresponds to model $f$, while
the dashed and dashed-dot lines correspond to models $n$10 and $n$12,
respectively. 

Figure (\ref{nparam_cen_a}a) shows the central
temperature (measured in ${\rm K}$) of the three models as a function of the central 
neutral density. At low densities, while the equation of state is
still isothermal, the central temperature remains constant and equal to
the $10 {\rm \, K}$ temperature of the reference state. In each model, at a density
equal to $n_{\rm opq}$, the equation of
state becomes adiabatic and the temperature begins to deviate from its
initial value. The evolution of temperature follows the
adiabatic law, $T \propto \rho^{\gamma - 1}$. The break in the power law and the 
subsequent, less steep increase, is the result of the excitation of the rotational
degrees of freedom of the neutral molecules, which corresponds to a
shift in the adiabatic index from $5/3$ to $7/5$. 
In the models we implement this change at a temperature of $200$ ${\rm
  K}$. The models do not all reach the same central density when each
run is stopped. This is because the termination criterion for each
run is that the central temperature reaches $1000$ ${\rm K}$, at which
thermal ionization (not accounted for in the model) 
 is expected to become important. This temperature is reached at a
 different central density in each model ($6 \times 10^{13}$, $6 \times 10^{14}$, 
and $6 \times 10^{15}$ ${\rm cm^{-3}}$ in models $n$10, $f$, and $n$12, respectively), 
depending only on the value of
 $n_{\rm opq}$.

Figure (\ref{nparam_cen_a}b) displays the dependence of the central
mass-to-flux ratio, normalized to its critical value, on the central
density of neutrals. The behavior in all cases is very similar to
that of the fiducial run (model $f$). 
The three curves deviate slightly from
one another close to the  density $n_{\rm opq}$ of each model, 
while they quickly asymptote to an identical behavior at 
higher $n_{\rm n,c}$. This is because $B_z$ exhibits only a mild 
dependence on the $n_{\rm opq}$ at high densities.

The evolution of the central magnetic field with density
(both $z-$ and $r-$ components, shown in Figs. \ref{nparam_cen_a}c
and \ref{nparam_cen_a}d, respectively) exhibits a relatively weak dependence on $n_{\rm opq}$. 
The most pronounced difference is caused by a steepening of the
density dependence of $B_{z,{\rm c}}$ immediately after the onset of
adiabaticity. If the equation of state had remained isothermal, 
$B_{z, {\rm c}}$ would have turned over and would have asymptoted to a constant value at high
densities (see Fig. \ref{iso_rad}c below). The behavior of $B_{z,{\rm c}}$ is similar 
in all three models {\em after} the onset of adiabaticity. However, 
in model $n$12 the overall value of $B_{z{\rm,c}}$ is
smaller at a given density than that in models $n$10 and $f$.
This is so because in model $n$12, adiabaticity is turned on at 
a density greater than that at which the profile of $B_{z,{\rm c}}$ begins to flatten, 
while in the other two models adiabaticity is turned on at a density smaller or about e
qual to that at which the flattening would begin. Despite this fact, all runs end at a
similar value of $B_{z,{\rm c}}\approx 10^{5} \mu {\rm G}$, because the final central 
density of each model increases with $n_{\rm opq}$.

The $r-$component of the magnetic field at the center of the 
contracting core $B_{r,{\rm c}}$ reaches a maximum at a 
density $n_{\rm n,c} \approx 10^{11}$ ${\rm  cm^{-3}}$ and decreases 
beyond that. This is because at that density the magnetic field begins 
to decouple from the matter and the field-line deformation can no longer
be sustained.

\begin{figure*}
\plotone{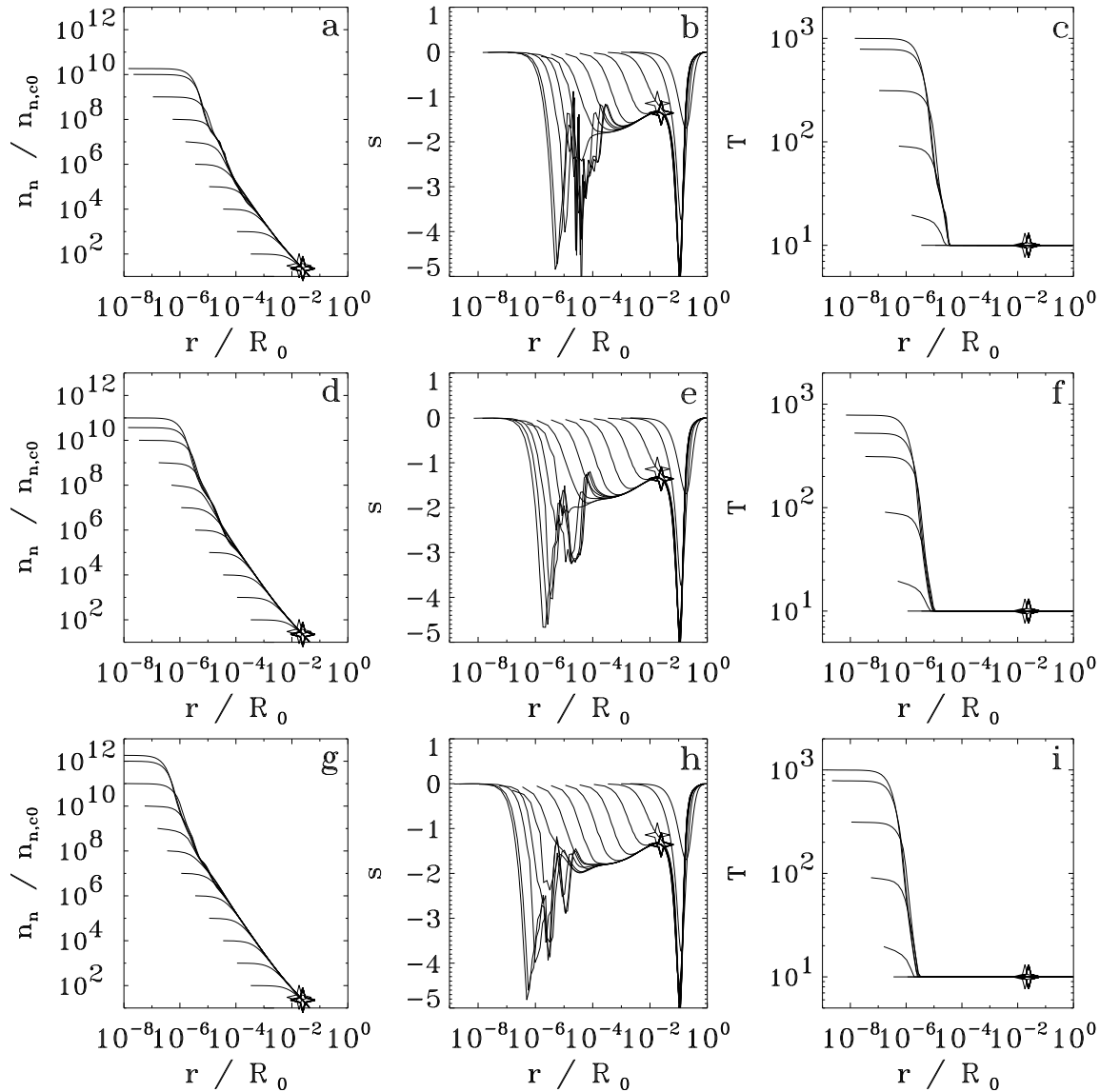}
\caption{\label{nparam_rad_a} 
Radial profiles, at different times, for models $n$10 (top row, $n_{\rm opq}=10^{10} 
{\, \rm cm^{-3}}$), $f$ (middle row, $n_{\rm opq}=10^{11} {\, \rm cm^{-3}}$), and $n$12 
(bottom row, $n_{\rm opq}=10^{12} {\, \rm cm^{-3}}$). The times are chosen such that 
neighboring curves differ by a factor of 10 in central neutral density. {\em Left column}: 
central number 
density of neutrals normalized to its initial value in the initial
equilibrium state ($n_{\rm n,c0}=2711$ ${\rm cm^{-3}}$); {\em middle column}: 
spatial derivative of the density $s=d\ln n_{\rm n}/ d\ln r$; {\em right column}: 
temperature (in ${\rm K}$).
} 
\end{figure*}

Figure (\ref{nparam_rad_a}) shows the radial dependence of the density
of the neutrals, $n_{\rm n}$, of its logarithmic 
derivative, $s=d\ln n_{\rm n}/dr$, and of the temperature, $T$, at different times. The top, 
middle and bottom rows refer to models $n$10, $f$, and 
$n$12, respectively.  Each curve corresponds to a central density enhancement of
one order of magnitude with respect to the previous curve,
except for the last curve, which always corresponds to a central
temperature of $1000 {\rm \, K}$. The ``star'' on a curve marks the location of the
supercritical core boundary.

The rightmost column of Figure \ref{nparam_rad_a} shows profiles of
the temperature. Only the five last curves in each case show deviation
from the reference-state temperature.  The radius at which 
adiabaticity sets in does not change appreciably with time. This is so because adiabaticity 
is implemented using a number-density criterion, and does
not change in time (see left column plots).
This radius is somewhat closer to the center the greater
$n_{\rm opq}$ is, because the density decreases monotonically with $r$.
The qualitative behavior of the number density of the
neutrals (left column) is the same as that discussed in Paper II (Fig. 3c) for
the fiducial run, with the power-law density profile steepening after
the onset of adiabaticity. The final central density achieved by the
time the temperature of $1000 {\, \rm K}$ has been reached increases with the value of 
$n_{\rm opq}$, and is therefore greatest in model $n$12. 

In Paper II (Fig. 3d) we discussed the qualitative behavior of the logarithmic
slope of the number density $s$ and pointed out that the two innermost dips are of thermal 
and magnetic origin, with the inner dip corresponding to a balance between the thermal-pressure 
and gravitational forces, and the outer dip corresponding to a
balance between magnetic and gravitational forces. The overall
magnitude of the thermal dip is essentially unchanged for different values of 
$n_{\rm opq}$. However, its location moves inward for greater $n_{\rm opq}$ as all of the 
post-adiabaticity features are translated toward smaller radii.
By contrast, the amplitude of the magnetic dip does
change with $n_{\rm opq}$; it is greater for smaller $n_{\rm opq}$.

\begin{figure*}
\plotone{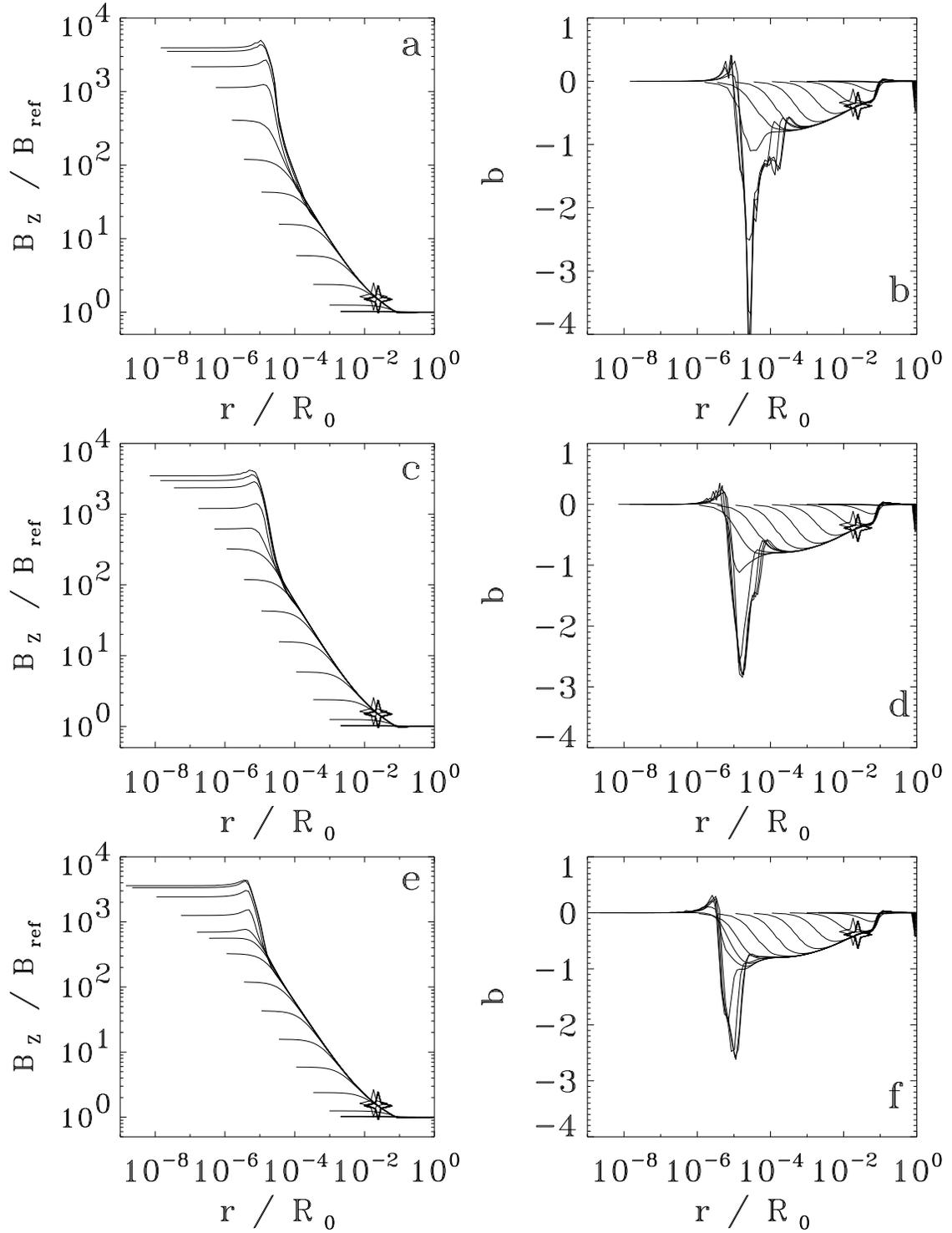}
\caption{\label{nparam_rad_b} 
Radial profiles of the $z-$component of the magnetic
field, normalized to the field strength of the reference state (left
column), and of its spatial derivative, $b=d\ln B_z/d\ln r$ (right
column) at different times, as in Figure (\ref{nparam_rad_a}). {\em Top row}: model 
$n$10; {\em middle row}: model $f$; {\em bottom row}: model $n$12.
}
\end{figure*}

Figure \ref{nparam_rad_b} shows radial profiles of the $z-$component of
the magnetic field (left column) and of its spatial derivative,
$b=d\ln B_z/d\ln r$ (right column). All models exhibit similar
behavior. There is a flat inner region followed by a local maximum at the boundary of the 
hydrostatic core. Just outside, the magnetic field decreases rapidly and the profile becomes
 steeper because the transition to adiabaticity occurs at smaller densities. In the isothermal
 regime there is a more gradual decrease with radius (mean slope
$\approx -0.7$) out to the boundary of the supercritical core. The profile of the magnetic 
field at the cloud envelope is not affected by the evolution of the supercritical core. 
Finally, the break in $b$ moves from $r/R_0 \approx 2 \times 10^{-5}$ in model $n$10 to 
$r/R_0 \approx 8 \times 10^{-6}$ in model $n$12, and the maximum value of $B_z$ from $r/R_0 
\approx 10^{-5}$ in model $n$10 to $r/R_0 \approx 5 \times 10^{-6}$ in model $n$12.

\begin{figure*}
\plotone{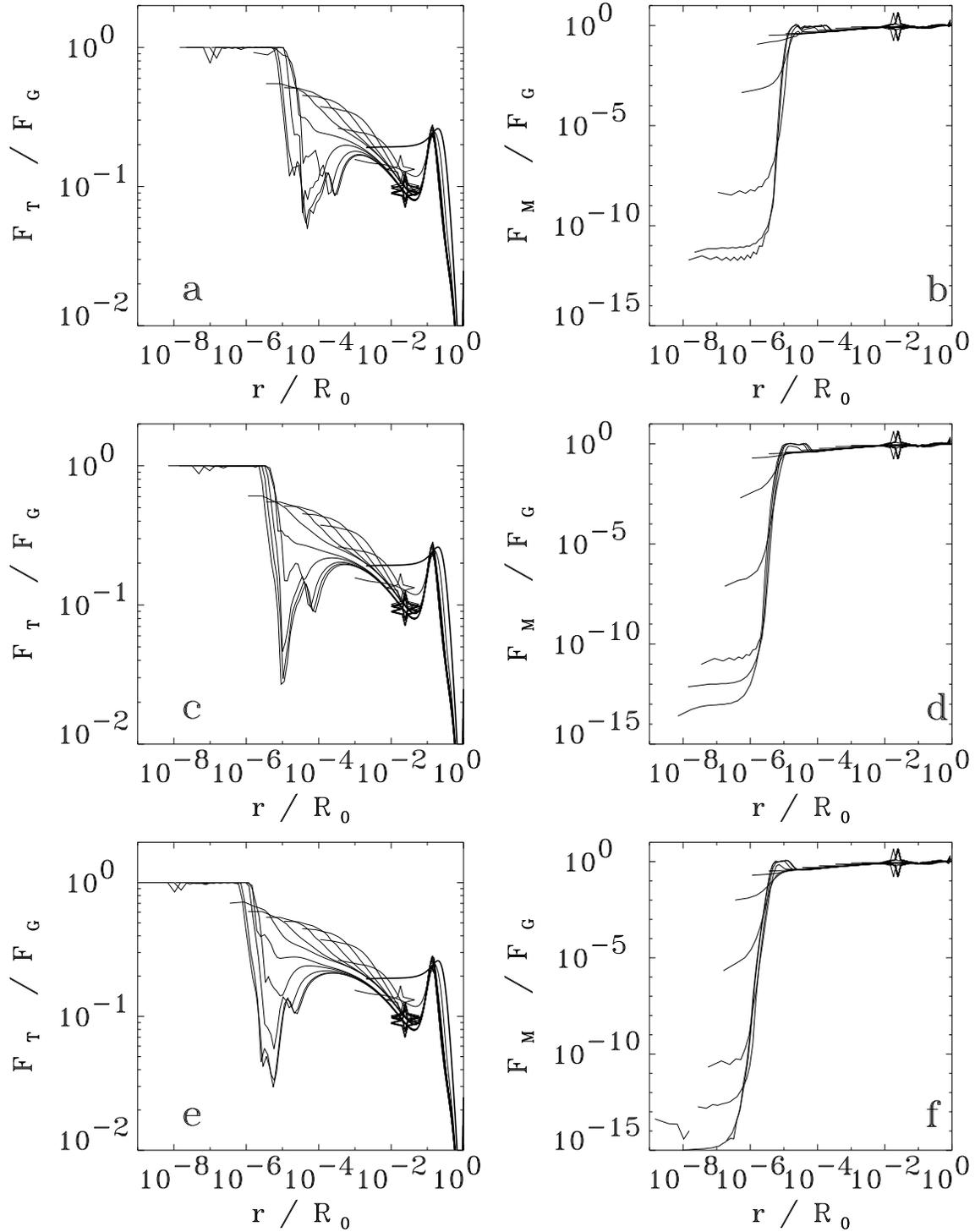}
\caption{\label{nparam_rad_c} 
Radial profiles of the ratio of thermal-pressure and gravitational forces (left
column), and magnetic and gravitational forces (right column) at different times,
chosen as in Figure (\ref{nparam_rad_a}).
{\em Top row}: model $n$10; {\em middle row}: model $f$; 
{\em bottom row}: model $n$12. The ``star'' marks the location of the supercritical core boundary.}
\end{figure*}

Figure \ref{nparam_rad_c} exhibits radial profiles of the ratio of
thermal-pressure and gravitational forces (left column) and magnetic and
gravitational force (right column) at different times. Each curve corresponds to
a central density enhancement of one order of magnitude with respect to the previous curve, 
except for the last curve, which always corresponds to a central temperature of 1000K. 
The thermal-pressure force becomes comparable to the gravitational force in the opaque core 
only after the onset of adiabaticity. The location of the thermal shock 
coincides with the hydrostatic core boundary. 
Because the density decreases monotonically with 
radius and the extent of the opaque region is limited by the condition $n\geq n_{\rm opq}$, 
the size of the hydrostatic core decreases with increasing $n_{\rm opq}$. 

The ratio of magnetic and gravitational forces exhibits the local
maximum discussed in Paper II, which is the cause 
of the magnetic
shock, and the corresponding magnetic feature in the density
profile. As $n_{\rm opq} $ increases, the location of this
maximum is displaced toward smaller radii, and the ratio
$F_{\rm M}/F_{\rm G}$ in the hydrostatic core reaches smaller values. This is so because, 
as $n_{\rm opq} $ increases, the transition from 
isothermality to adiabaticity takes place at higher densities and the 
termination condition for each run ($T = 1000$ $K$) is achieved at 
greater central densities.


The small-amplitude oscillations which appear in the ratio of 
 magnetic and gravitational forces inside the hydrostatic core 
are numerical noise due teh unavoidable inaccuracies 
involved in calculating spatial derivatives (needed to derive 
the magnetic force) inside the hydrostatic core, where the 
magnetic field is very nearly spatially uniform. However, 
because in this region the magnetic field is dynamically 
unimportant, the impact of such noise on our results is 
insignificant. The disconnected line in panel (f) is due to logarithmic 
plotting (in the region where the line has not been plotted, the value of
the magnetic force has become numerically zero).


\begin{figure*}
\plotone{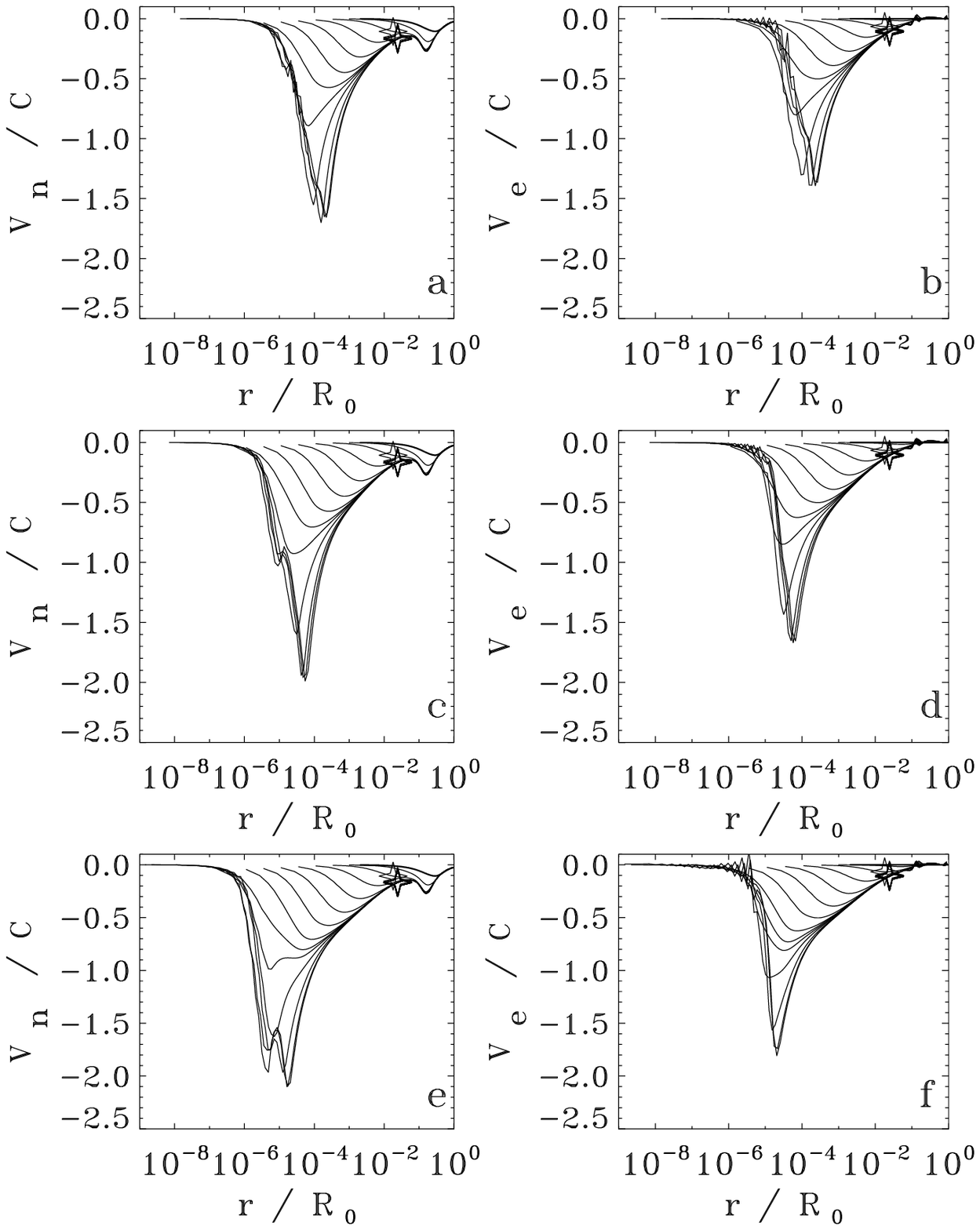}
\caption{\label{n_param_rad_d}  Spatial profiles of the radial velocities
of the neutrals (left column) and the electrons (right column), normalized to 
the sound speed of the reference state, at different times chosen as in Figure 
(\ref{nparam_rad_a}). {\em Top row}: model $n$10; {\em middle row}: model $f$; 
{\em bottom row}: model $n$12. The 
``star'' marks the location of the supercritical core boundary.}
\end{figure*}

Figure \ref{n_param_rad_d} shows the spatial profiles of the radial velocities
of the neutrals (left column) and the electrons (right column) in
units of the sound speed (in the neutrals) of the reference state, for the three
models $n$10, $f$ and $n$12 (top to bottom row). 
All models exhibit the
same qualitative behavior: There are always two features in the velocity of the neutrals, 
the ``thermal'' shock and the ``magnetic'' shock at somewhat greater radii. The velocity 
of the electrons (which are, except at very small radii in
model $n$12, well attached to the magnetic field)  exhibits only the magnetic

shock. Its effect is much more pronounced than in the case of the
neutrals, and the electrons decelerate almost to a halt, at the boundary of the
hydrostatic core.

The positions of the magnetic and the thermal shocks move to
smaller radii with increasing $n_{\rm opq}$, because the boundary of the
hydrostatic core (which moves inward with increasing $n_{\rm opq}$)
determines the position of the thermal shock and, by extension, the
position of the magnetic feature, which is caused by field lines
piling up {\em just outside} the hydrostatic core. 
The maximum infall velocity of both neutrals and electrons increases somewhat
as $n_{\rm opq}$ increases, because the delay of the onset of
adiabaticity implies accelerated infall for longer times. In addition,
the reacceleration of the neutrals between the magnetic and the
thermal shocks is greater with increasing $n_{\rm opq}$. This is because, as
seen in Figure \ref{nparam_rad_c}, in the case of smaller $n_{\rm
  opq}$ the thermal-pressure force as a fraction of the gravitational force
increases faster with decreasing radius between the magnetic shock and
the hydrostatic core, and hence the reacceleration is milder.


\begin{figure*}
\plotone{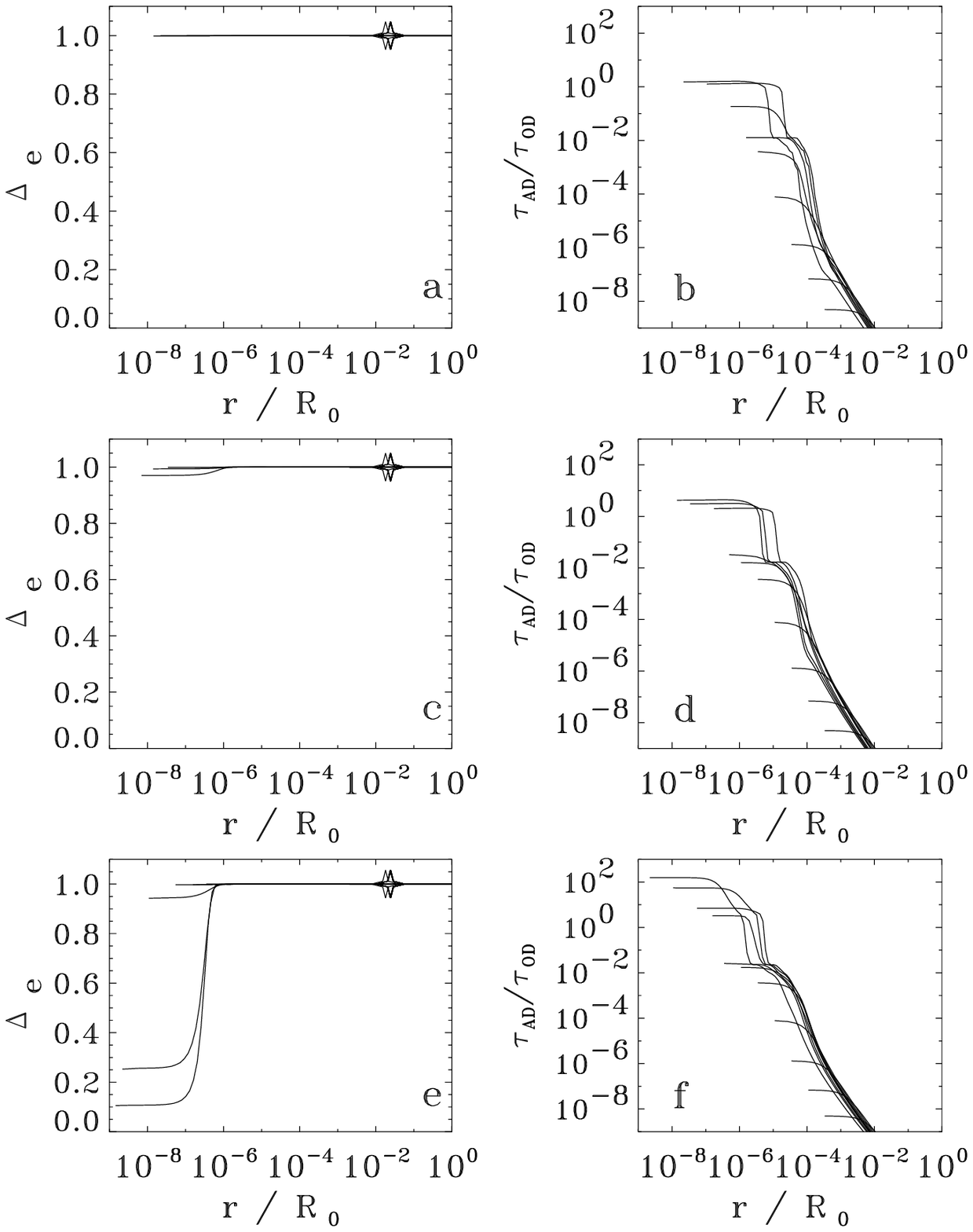}
\caption{\label{nparam_rad_e} 
Radial profiles of the attachment parameter of the electrons
(left column) and the ratio of ambipolar-diffusion and Ohmic-dissipation 
timescales (right column) at different times, chosen as in Figure (\ref{nparam_rad_a}). 
{\em Top row}: model $n$10; {\em middle row}: model $f$; {\em bottom row}: model $n$12. The 
``star'' marks the location of the supercritical core boundary.}
\end{figure*}

The left column of Figure \ref{nparam_rad_e} provides crucial
information concerning the importance of the magnetic field during the
late stages of star formation and whether a complete detachment of the
field from the matter can occur before thermal ionization
reattaches matter to the field lines at $T \approx 1000$ K. At the late stages of
contraction of the supercritical core, which are for the first time
followed in this series of papers, ions have detached from the magnetic field lines
and follow the motion of the neutrals. Hence, the electrons are the only particles
that may remain attached to the field lines until the onset of thermal
reionization of the core. The degree of attachment of electrons to the field lines 
is measured, as we have discussed in Papers I and II, by the attachment
parameter $\Delta_{\rm e}$. The value $\Delta_{\rm e} \approx 1$
indicates that the electrons are well coupled to the field lines,
while $\Delta_{\rm e} \approx 0$ indicates that the electrons have
detached and follow the motion of the neutrals. 

As seen in the left-column plots of Figure \ref{nparam_rad_e},
in the model with $n_{\rm opq} =10^{10} {\, \rm cm ^{-3}}$, the electrons remain
well attached to the magnetic field throughout the run. In the case 
$n_{\rm opq} =10^{11} {\, \rm cm ^{-3}}$, the electrons begin to
detach before the central temperature reaches $1000 {\rm \, K}$, at which 
the run is stopped. However, their degree of attachment is still
high. Finally, in the model with $n_{\rm opq} =10^{12} {\, \rm cm ^{-3}}$, the
electrons in the core have detached almost completely before the end of the run 
($\Delta_{\rm e} \approx 0.1$). 

Using the indirect attachment parameter of the electrons 
to trace the densities at which the electron detachment begins and comes to completion, 
we find that electrons begin to detach at about $10^{14}$ ${\rm cm^3}$ and detach completely 
by a density $\approx 10^{16}$ ${\rm cm^{-3}}$, regardless of the value of $n_{\rm opq}$. 
Hence, we arrive at the conclusion that {\em the
detachment of electrons  (and consequently the complete decoupling of
matter) from the magnetic field lines takes place at a number density $\approx
10^{15}$ ${\rm cm^3}$, provided that this density can be reached before thermal ionization 
becomes important}. In order to confidently assert whether the required density is achieved 
before thermal ionization sets in, and hence whether this 
complete decoupling of matter from the magnetic field occurs in nature, the temperature 
history of the collapsing core would need to be followed in detail,
accounting for radiative transfer. However, given the results of 
our parameter study, the detachment of electrons before the onset of
thermal ionization is unlikely. This is so because the actual temperature will increase 
more by the time the central density becomes $10^{15}$ ${\rm cm^3}$ than it does in the isothermal
approximation up to $10^{12} {\, \rm cm^{-3}}$, beyond which adiabaticity was assumed.

A second important issue during the very late stages of collapse is
whether and when Ohmic dissipation overtakes ambipolar diffusion as
the dominant flux reduction mechanism. This issue is explored
by the right-column plots of Figure \ref{nparam_rad_e}, which show
radial profiles of the ratio of the ambipolar-diffusion and 
Ohmic-dissipation timescales. The ambipolar-diffusion timescale is 
smaller at all radii and
times, except in the very late stages of the contraction,
at densities $\gtrsim 10^{13} {\rm \, cm^{-3}}$. For densities $\sim
10^{16} {\,\rm cm^{-3}}$, the Ohmic-dissipation timescale becomes $\approx
100$ times smaller than the corresponding ambipolar-diffusion
timescale. {\em Thus the density at which Ohmic dissipation becomes 
more effective than ambipolar diffusion is not sensitive to the 
thermal history of the protostellar core, but the size of the region in  
which this happens is: The greater the density at which the protostellar core becomes 
opaque, the smaller the region in which Ohmic dissipation dominates}.

\subsection{The Isothermal Control Run}
\label{iso}

In this section we present results of a ``control run'' for which the
assumption of isothermality is retained throughout the evolution of the
 contracting model cloud. These results
complement the parameter study with regard to $n_{\rm opq}$, by
allowing one to identify the features that can be attributed to the
transition between equations of state. All parameters other than the
equation of state are identical with those of the runs discussed in the
previous section.

\begin{figure*}
\plotone{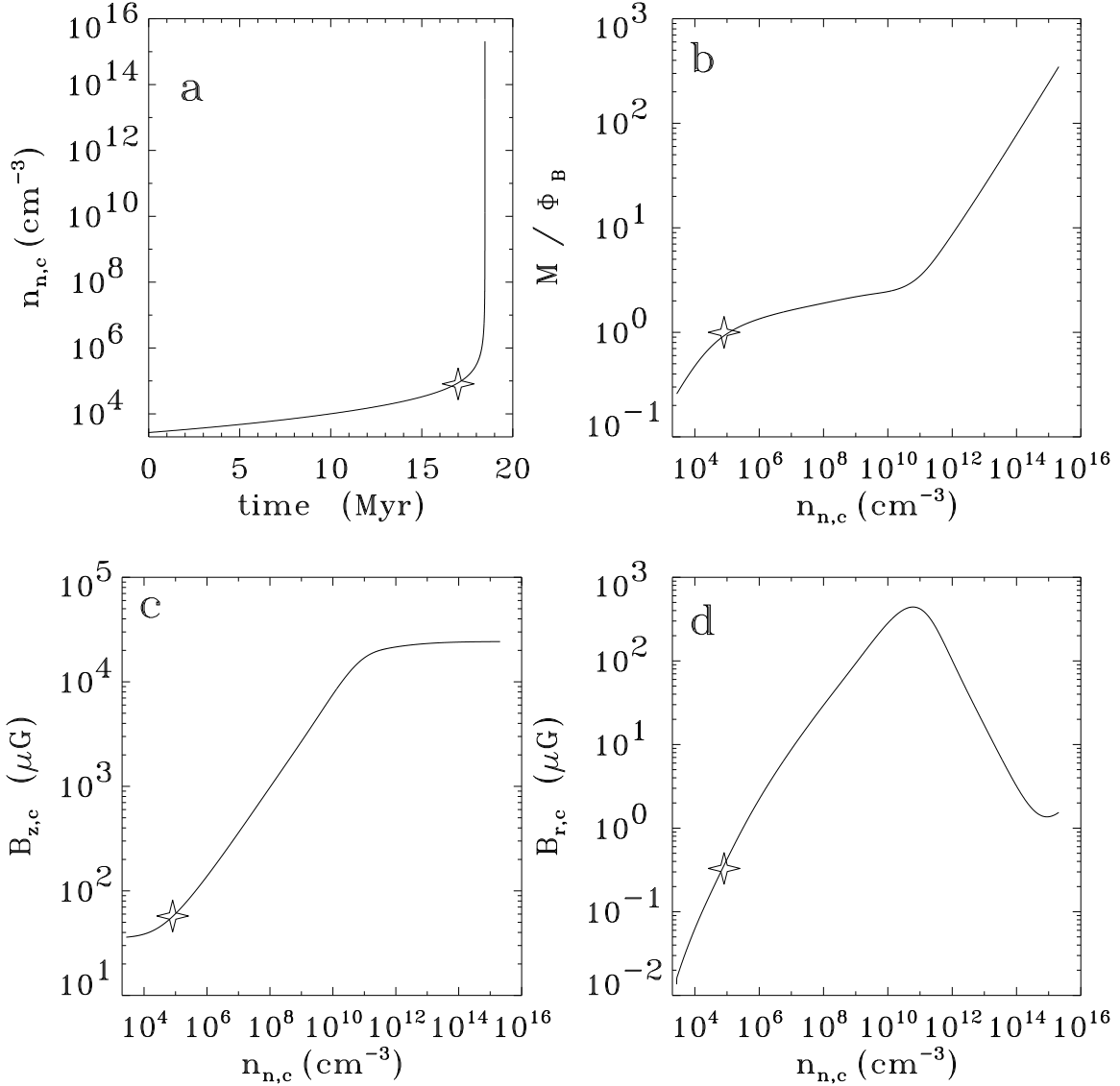}
\caption{\label{iso_rad} 
Evolution of central quantities for model $i$ (isothermality assumption
retained throughout the run). (a) Central number density of neutrals, normalized
to its initial value in the reference state, as a function
of time (in Myr); (b) mass-to-flux ratio, in units of the critical value,
as a function of the central density enhancement; (c) $z-$component
of the magnetic field, normalized to the field strength of the
reference state, as a function of the central density enhancement;
(d) $r-$component of the magnetic field, normalized to the field strength of the
reference state, as a function of the central density enhancement.
The ``star'' marks the formation of the supercritical core.}
\end{figure*}

Figure (\ref{iso_rad}) shows the evolution of central quantities for this model. 
The dependence of the central density of neutral particles on time
is plotted in Figure (\ref{iso_rad}a). This curve is essentially identical to the
corresponding one of the fiducial run discussed in Paper II, since
the slow, ambipolar-diffusion--controlled early stage of the formation
of the supercritical core occurs isothermally in both models. 

The evolution of the central mass-to-flux ratio and of the $r$-component of
the magnetic field at the center of the cloud, shown in Figures
(\ref{iso_rad}b) and (\ref{iso_rad}d), respectively, remained essentially the same for
different values of $n_{\rm opq}$, as we saw in the previous section.
It is also very similar in the isothermal run. However, the
evolution of the $z$-component of the magnetic field, shown in Figure
(\ref{iso_rad}c), shows a qualitatively different behavior at late
times. In the previous section it was shown that, after adiabaticity
sets in, the dependence of $B_{z,{\rm c}}$ on density steepens, and
$B_{z,c}$ keeps increasing until the end of the run. In the isothermal
model, this steepening never takes place. Instead, the
curve flattens and $B_{z,{\rm c}}$ asymptotically approaches $0.28$ ${\rm G}$, a
value which is about $800$ times the magnetic field of the initial 
reference state, and slightly greater than the one reached in the 
adiabatic runs by the time the temperature became 1000 K.

\begin{figure*}
\plotone{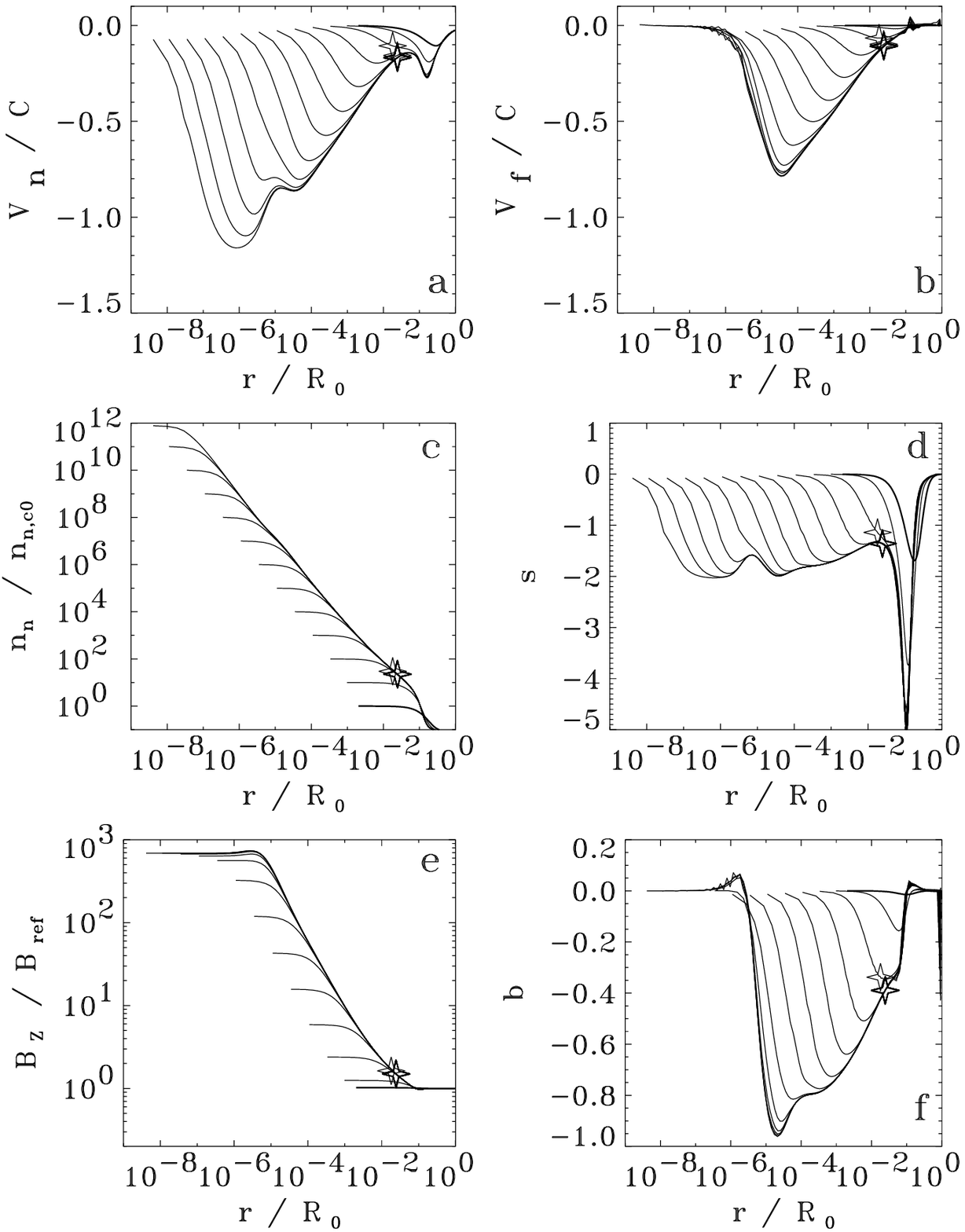}
\caption{\label{iso_rad_b} 
Spatial profiles of physical quantities of model $i$ (isothermal run) at different 
times, chosen as in Figure (\ref{nparam_rad_a}). (a) Velocity of neutrals normalized to
the sound speed of the reference state; (b) velocity of the
field lines normalized to the sound speed of the reference state; 
(c) number density of the neutrals normalized to the initial central
density of the reference state; (d) slope of the density
profile $s=d\ln n_{\rm n}/ d\ln r$; (e) $z-$component of the  
magnetic field normalized to the field strength of the reference
state; (f) slope of $B_z$, $b=d\ln B_z/d\ln r$. The ``star''
marks the location of the supercritical core boundary.}
\end{figure*}

Figures (\ref{iso_rad_b}c) and (\ref{iso_rad_b}d) show the neutral number density profile
and its derivative, $s=d\ln n_{\rm n} /d\ln r$, respectively, at
different times chosen as in Figure (\ref{nparam_rad_a}). As before, each curve corresponds 
to an order of magnitude enhancement in the central density relative to the previous curve. 
In \S \, \ref{nop}, (in relation to Fig. \ref{nparam_rad_a}) and in Paper II, we identified 
two types of features inside the supercritical core but outside the innermost uniform-density
region in the density profile and its derivative: an inner,
``thermal'' steepening of the profile, corresponding to a dip in the slope
$s$, and a second, ``magnetic'' steepening, due to the local maximum
in the $z-$component of the magnetic field and the corresponding
balance between magnetic and gravitational forces. In the isothermal
run, the inner thermal feature is completely
absent, because the temperature remains always at $10 {\, \rm K}$ and a
hydrostatic core never forms. The magnetic feature is still
present, but is less pronounced because the enhanced ``pile-up'' of
matter and magnetic flux outside the hydrostatic core
is now absent. 

Spatial profiles of the velocities of the neutral molecules and the field lines are shown in 
Figure (\ref{iso_rad_b}a) and (\ref{iso_rad_b}b), respectively. Again,
only the magnetic feature is present here, at $r=10^{-5} R_0$, and it is much
weaker than the corresponding magnetic shocks seen in the runs that
allow adiabaticity at the highest densities. The reason for this
difference is that the presence of the hydrostatic core leads to a
more pronounced accumulation of flux just outside its boundary.


Figures (\ref{iso_rad_b}e) and (\ref{iso_rad_b}f) display spatial profiles of $B_{z}$ and 
its logarithmic derivative, $b$, at different times as in Figure (\ref{nparam_rad_a}). 
Although the qualitative behavior of $B_z$ is similar to that of the models discussed above 
(moving from larger to smaller radii, a steepening of the power-law profile, followed by a 
local maximum and then the innermost flat region), all features are less pronounced in the 
present, isothermal case. This is especially evident in the case of the local maximum,
because of the ``pile-up'' of field lines and matter that occurs just outside the hydrostatic 
core in the adiabatic case. 

\begin{figure*}
\plotone{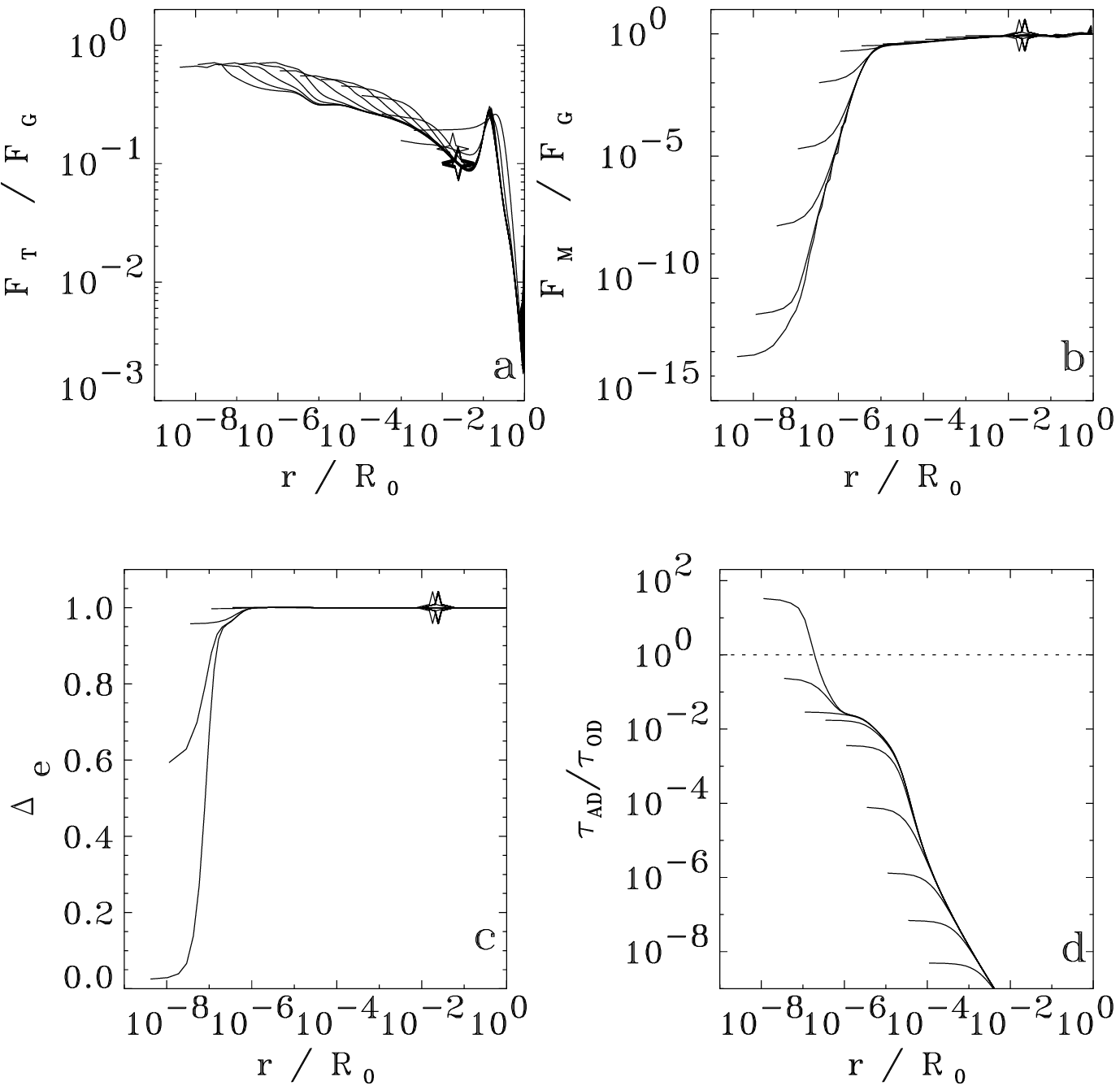}
\caption{\label{iso_rad_c} 
Radial profiles of physical quantities of model $i$ (isothermal 
run) at different times, chosen as in Figure (\ref{nparam_rad_a}). (a) Ratio of 
thermal-pressure and gravitational forces; (b) ratio of magnetic and gravitational 
forces; (c) attachment parameter of electrons; (d) ratio of ambipolar-diffusion
and Ohmic-dissipation timescales (the dotted line corresponds to
$\tau_{\rm AD} = \tau_{\rm OD}$). The ``star'' marks the location of the
supercritical core boundary.}
\end{figure*}

Figures (\ref{iso_rad_c}a) and (\ref{iso_rad_c}b) show the spatial variation of the ratios
of thermal-pressure and gravitational forces, and magnetic and
gravitational forces, respectively. In the isothermal model, the
thermal-pressure force never balances the gravitational force. However, in the
innermost part of the collapsing supercritical core at late
times, the thermal-pressure force overtakes the magnetic force. In this region, then, at late 
times, the thermal-pressure forces provide the main opposition to gravity even 
in this isothermal run.

The spatial variation of the attachment parameter of the electrons, $\Delta_{\rm e}$, is shown 
in Figure (\ref{iso_rad_c}c) at different times, as in Figure (\ref{nparam_rad_a}). As discussed 
above, the detachment of the electrons is primarily
density related. The electrons are completely detached from the field lines by $n_{\rm n} \approx 
10^{16}$ ${\rm cm^{-3}}$. 

The ratio of the ambipolar-diffusion and Ohmic-dissipation timescales is shown in Figure 
\ref{iso_rad_c}d as function of radius, normalized as usual to the initial cloud radius $R_0$, 
at different times. Below $n_{\rm n,c} \approx 10^{14}$ ${\rm cm^3}$,
ambipolar diffusion is much more effective than Ohmic dissipation in
increasing the mass-to-flux ratio in any given flux tube. It is
only for densities greater than $\approx 10^{14}$ ${\rm cm^3}$ that 
Ohmic dissipation becomes more important than ambipolar diffusion. 
This density is slightly greater than in the adiabatic cases.


Having examined the limiting isothermal case, we conclude that the results concerning the 
structure of the magnetic field, the evolution of the ratio of 
ambipolar-diffusion and Ohmic-dissipation timescales, and the density 
at which the electrons detach from the magnetic field, are not sensitive 
to the assumed equation of state. However, the presence of a hydrostatic
core in the adiabatic models accelerates the formation of the ``magnetic wall''
(the evolution of which eventually leads to the series of magnetically
driven shocks studied in Tassis \& Mouschovias 2005a,b).


\section{Dependence on the Mass-to-Flux Ratio}\label{mtfvar}

\begin{figure*}
\plotone{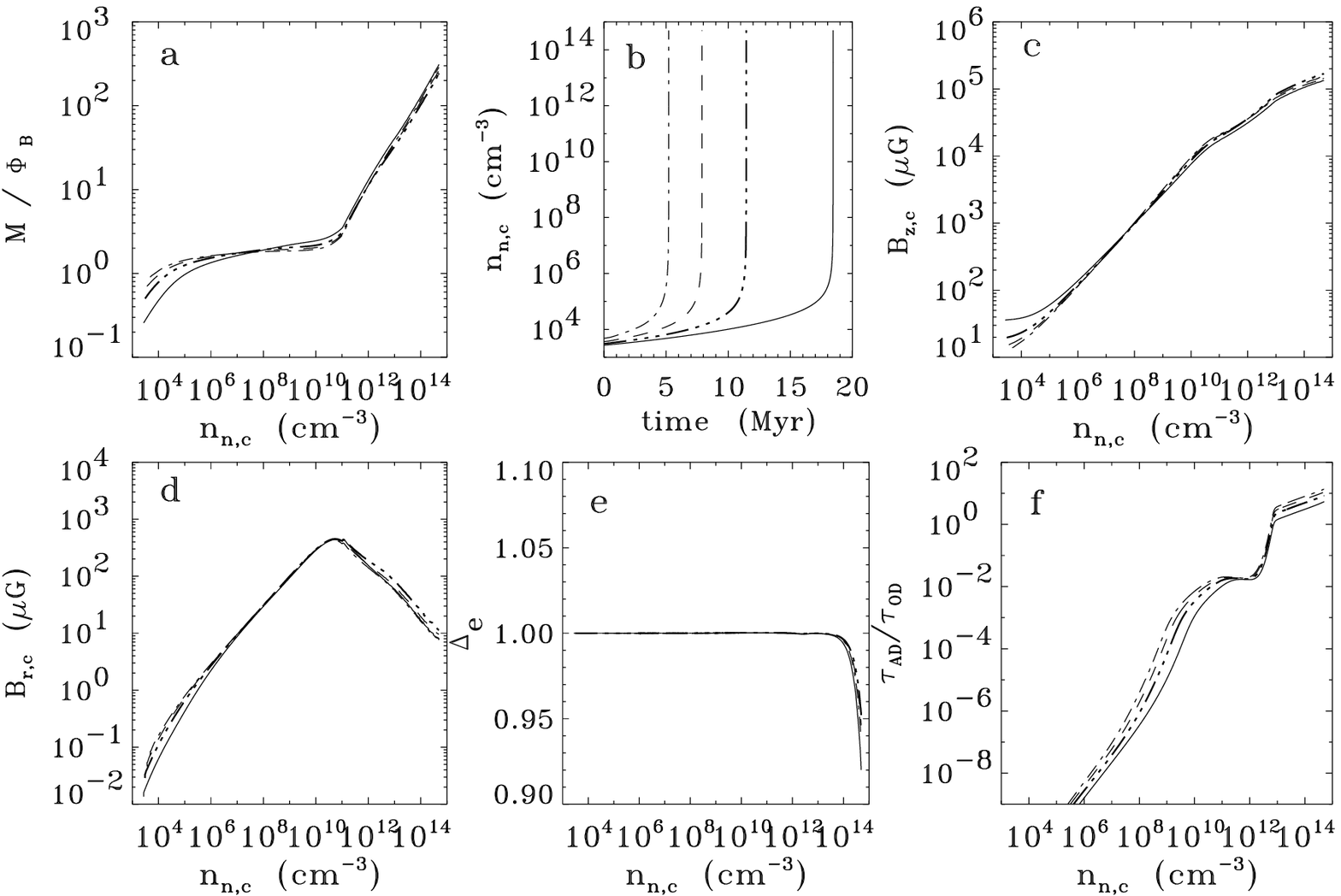}
\caption{\label{mfparam_cent_a} 
Evolution of central quantities for models $f$ (solid line, $\mu_{\rm d,c0} = 0.25$), 
$m$5 (triple-dot--dashed line,  $\mu_{\rm d,c0} = 0.5$), $m$7  (dashed line, $\mu_{\rm d,c0} 
= 0.7$) and $m$9 (dashed-dot line,  $\mu_{\rm d,c0} = 0.9$).  (a) Central mass-to-flux ratio, 
normalized to the critical value, as a function of the central number density of
neutrals; (b) central number density of neutrals as a function of time;
(c) and (d) $z-$ and $r-$components, respectively, of the magnetic field, as
functions of neutral density; (e) attachment parameter of the electrons; (f) ratio of 
ambipolar-diffusion and Ohmic-dissipation timescales. 
}
\end{figure*}

\begin{figure*}
\plotone{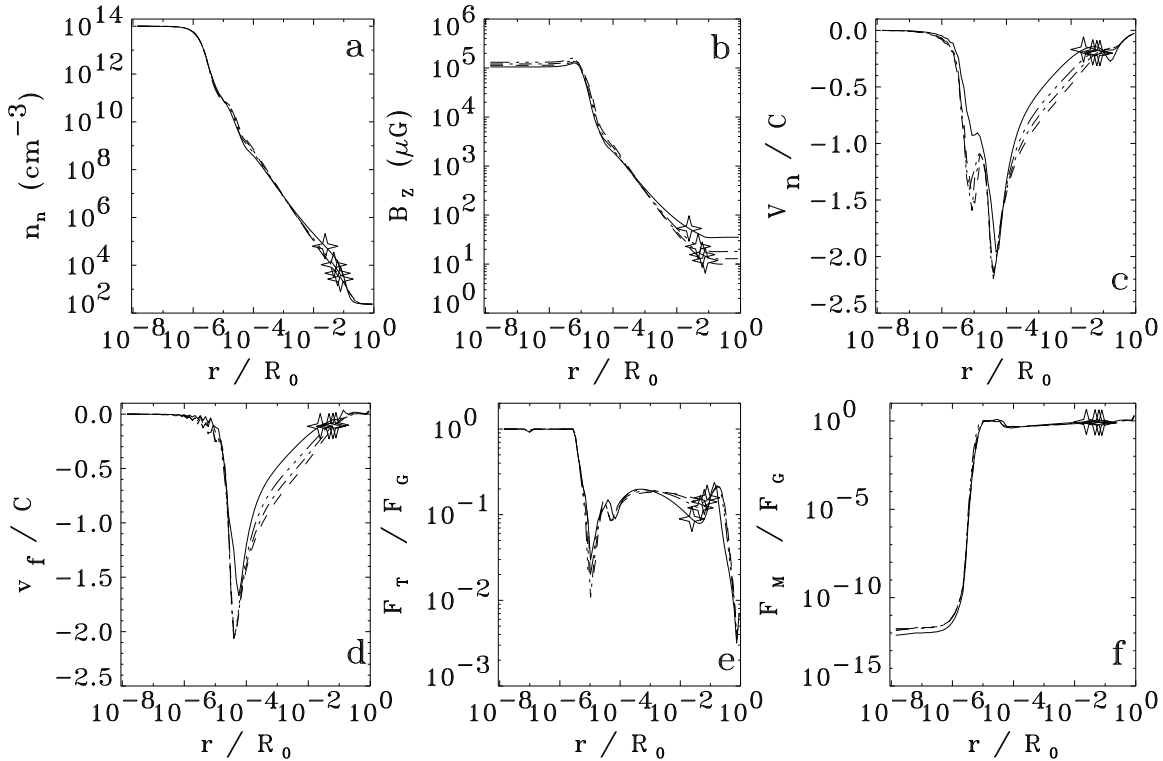}
\caption{\label{mfparam_rad_a} Radial profiles for models $f$ (solid line),  $m$5 
(triple-dot--dashed line), $m$7  (dashed  line) and $m$9 (dashed-dot line), when the 
central density is $n_{\rm n,c} = 10^{14} \, \, {\rm cm^{-3}}$. (a) Central density of 
neutrals; (b) $z-$component of the magnetic field; (c) radial velocity of neutrals;
(d) radial velocity of magnetic field lines; (e) ratio of thermal-pressure and
gravitational forces; (f) ratio of magnetic and gravitational
forces. The ``star'' on each curve marks the location of the supercritical core boundary.}
\end{figure*}

In order to quantify the effect of the initial mass-to-flux
ratio of the parent cloud on the evolution of the supercritical
core, we compare four models ($f$, $m$5, $m$7 and $m$9). 
All have the same value of $n_{\rm opq}=10^{11} {\, \rm cm^{-3}}$, but an
initial central mass-to-flux ratio (in units of the
critical value for collapse) equal to $0.25$, $0.5$, $0.7$ and $0.9$,
respectively. In Figures (\ref{mfparam_cent_a}) and (\ref{mfparam_rad_a}),
model $f$ is represented by a solid line, model $m$5 by a
triple-dot--dashed line, model $m$7 by a dashed line, and model $m$9
by a dashed-dot line. 

Figure (\ref{mfparam_cent_a}a) shows the  central mass-to-flux ratio in
units of the critical value as a function of the central number
density of neutrals. Although the initial mass-to-flux ratio 
is different in these models, {\em all memory of the initial conditions is lost after the 
formation of the supercritical core , and the increase of the central mass-to-flux ratio 
with density is practically identical in all models}. A very similar 
behavior is exhibited by the ($z-$ and $r-$) components of the central magnetic
field, shown as functions of central density in Figures \ref{mfparam_cent_a}c and 
\ref{mfparam_cent_a}d, respectively. 
The final values of $B_z$ (and $B_r$) in all models differ by less than a factor of
2. This behavior is even more pronounced in the case of $\Delta_{\rm e}$, the attachment 
parameter of the electrons, which is shown as a function of the central number density 
of neutrals in Figure \ref{mfparam_cent_a}e. $\Delta_{\rm e}$ starts out
at the same value in all models ($\Delta_{\rm e}=1$, which corresponds
to perfect attachment of the electrons to the field lines), and
evolves in the same manner in all models. At the end of each run, by the time a central 
temperature of $1000 {\rm \, K}$ is reached, its values in the four models are: 0.92 
(model $f$), 0.94 (model $m$5), and 0.95 (models $m$7 and $m$9).


The quantity that {\em is} significantly affected by the initial
mass-to-flux ratio is the time required for a particular central
density to be reached (see Fig. \ref{mfparam_cent_a}b). 
The cloud spends most of the time in the early phase, until a supercritical core forms. 
Thereafter the contraction becomes dynamic (accelerated, but much slower than free fall). 
The total evolution time, from the initial state until the end of the run, ranges from 5 
Myr (model $m$9) to 17 Myr (model $f$).

Figure (\ref{mfparam_cent_a}f) shows the ratio of the ambipolar-diffusion 
and Ohmic-dissipation timescales at the center of the cloud, as a function of the central 
number density of neutral particles. Although the quantitative differences among models
before the onset of adiabaticity are appreciable, the qualitative behavior is similar in 
all models. Models with smaller initial mass-to-flux ratio exhibit systematically smaller
$\tau_{\rm AD}/\tau_{\rm OD}$ values before the onset of adiabaticity.
After adiabaticity sets in, the evolution of the timescales ratio is similar in all models. 
Ambipolar diffusion dominates Ohmic dissipation early on, but the two mechanisms become
equally effective at central densities $\approx 10^{13} {\, \rm cm^{-3}}$ in all cases. 

Figure (\ref{mfparam_rad_a}) shows radial profiles of different
quantities for models $f$, $m$5, $m$7 and $m$9. All curves correspond to that
instant in time when the central number density of neutrals reaches
the value $n_{\rm n,c} = 10^{14} {\, \rm cm^{-3}}$. On each curve in these plots, 
the ``star'' marks the location of the boundary of the supercritical core in the corresponding model.

Figure (\ref{mfparam_rad_a}a) shows the central number density of
neutrals as a function of radius for each of the four models. At the late time
corresponding to these curves, the established density profile is
practically identical in all models, except near the boundary of the supercritical 
core, where small deviations between models can be identified. Just outside the 
boundary of the supercritical core, the density profile is steeper for smaller 
mass-to-flux ratios, because the envelopes are better supported by magnetic forces.

Figure (\ref{mfparam_rad_a}b) shows the profiles of the $z-$component of
the magnetic field for the four models. The field strength
in the initial state of the parent cloud, which is preserved in
the outermost regions of the model cloud, is different in each
model, and is greater in models with smaller mass-to-flux
ratio. However, within one order of magnitude in radius inward from the
location of the supercritical core, the profile of $B_{z}$ is almost the same
in all models.


Figure (\ref{mfparam_rad_a}c) shows profiles of the radial velocity of
the neutral particles in units of the sound speed of the reference
state. Although the behavior is qualitatively the same for all models,
the evolution of the velocities and the values at their local
minima depend on the initial mass-to-flux ratio. Inward of the
location of the supercritical core boundary, the
magnitude of the infall velocities achieved at a certain radius is
greater for smaller initial mass-to-flux ratio. The reason for this
trend is that the dynamical contraction sets in at a greater central density
enhancement as the initial mass-to-flux ratio decreases. As a
result, by any given central number density, the core 
will have spent less time in the dynamical phase in the models with smaller mass-to-flux
ratio, and the resulting velocities will be smaller at any given
radius. For this reason, the maximum infall velocity
achieved by the neutrals increases as the initial mass-to-flux
ratio increases, and is greatest in model $m$9.

Inward of the magnetic shock, the neutrals are reaccelerated by gravity before the
thermal shock, behind which they almost come to rest in the hydrostatic core. The 
reacceleration is smaller for smaller initial
mass-to-flux ratios because, as we have seen, the density profile is somewhat steeper
at radii smaller than the radius at which adiabaticity sets in; hence, the
thermal-pressure force is greater, presenting a greater opposition to gravity.

This interpretation is verified by Figure (\ref{mfparam_rad_a}e), which shows
profiles of the ratio of thermal-pressure and gravitational forces in each
model. The thermal-pressure force is indeed a greater fraction of
the gravitational force in the reacceleration region (inward of the
magnetic shock) in the models with smaller mass-to-flux ratios. The magnetic force, on
the other hand, decreases rapidly with decreasing radius inward of
the plateau of the magnetic force, where it balances gravity. In
addition, the magnetic force as a fraction of the gravitational force
has almost identical behavior in all models at a given value of the
density. This can be seen  clearly in Figure (\ref{mfparam_rad_a}f), which plots the 
ratio of the magnetic and gravitational force in each model. 

The profiles of the velocity of the field lines, $v_{\rm f}$,
 in the four models are compared in Figure (\ref{mfparam_rad_a}d). The velocity of the
field lines exhibits only the magnetic-wall feature, because the deceleration
at the magnetic shock brings them, as well as the attached species (the electrons), 
almost to rest. Hence the presence of the hydrostatic core boundary does not affect the motion
of the field lines significantly. The maximum infall velocity exhibited by the field
lines is greater in models with greater initial mass-to-flux ratios, because
dynamical contraction operates for a longer time.

\section{$^{26}$Al as an Alternative Radionuclide}\label{alvar}

All models discussed above assumed that, at very high densities, when the
central part of the collapsing core becomes self-shielded from cosmic-ray
ionization, the ionization rate is dominated by 
$^{40}$K radioactivity (half-life = 1.25 Gyr). If, however, the
initial mass-to-flux ratio of the parent cloud is relatively close to critical,
then the evolution is rapid and radionuclides with smaller lifetimes (such as $^{26}$Al, 
with a half-life = 0.716 Myr) can dominate the ionization at high densities. Similarly, 
$^{26}$Al can become important if the core happens to get enriched because of a nearby 
Supernova explosion. In this section we compare results from models $m$9 and $Al$, both 
of which have an initial central mass-to-flux ratio equal to $0.9$ times the critical 
value and $n_{\rm opq} = 10^{11}$ ${\rm
  cm^{-3}}$. Their difference lies in the radionuclide assumed to
dominate high-density ionization, which is $^{40}K$ in model $m$9 and
$^{26}Al$ in model $Al$. 


\begin{figure*}
\plotone{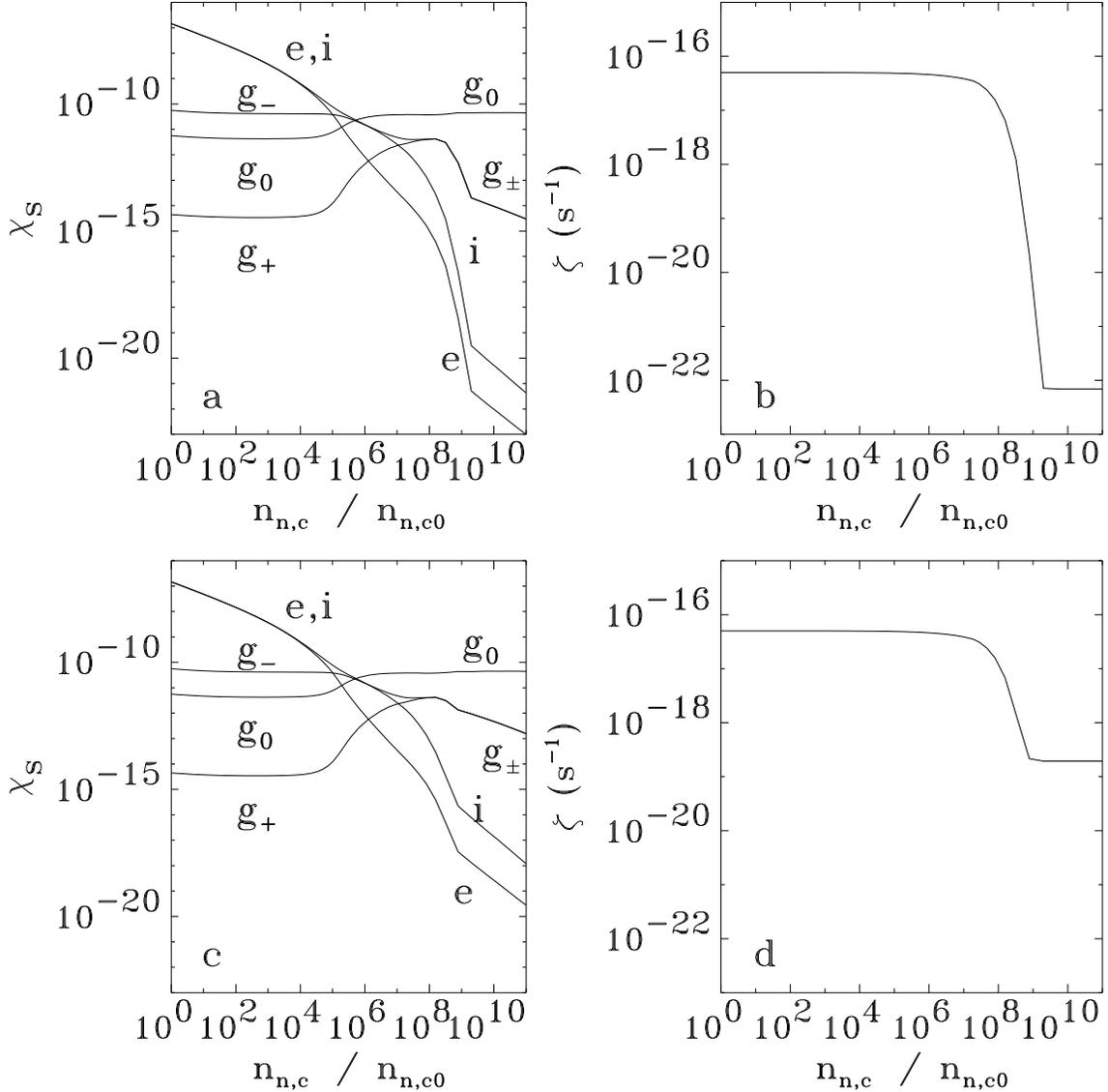}
\caption{\label{al_a} 
Abundances of different species (left column) and ionization rate
(right column) as functions of the central neutral density normalized to its initial value 
($n_{\rm n,c0} = 4900 \, {\rm cm^{-3}}$).
{\em Top row}: model $m$9 ($^{40}$K is the dominant radionuclide); {\em bottom
row}: model $Al$ ($^{26}$Al is the dominant radionuclide).} 
\end{figure*}

Figure (\ref{al_a}) shows chemical abundances of different species as
fractions of the number density of neutral hydrogen molecules (left
column) and the ionization rate as a function of number density (right
column) for models $m$9 (top row) and $Al$ (bottom row). 
In the case of model $Al$, the plateau reached by the ionization rate at high densities 
occurs at considerably greater values (more than three orders of magnitude) than in 
model $m$9. As a result, the degree of ionization and the 
chemical abundances of all charged species are significantly greater at
high densities, when the ionization rate is dominated by radioactive
decays.

\begin{figure*}
\plotone{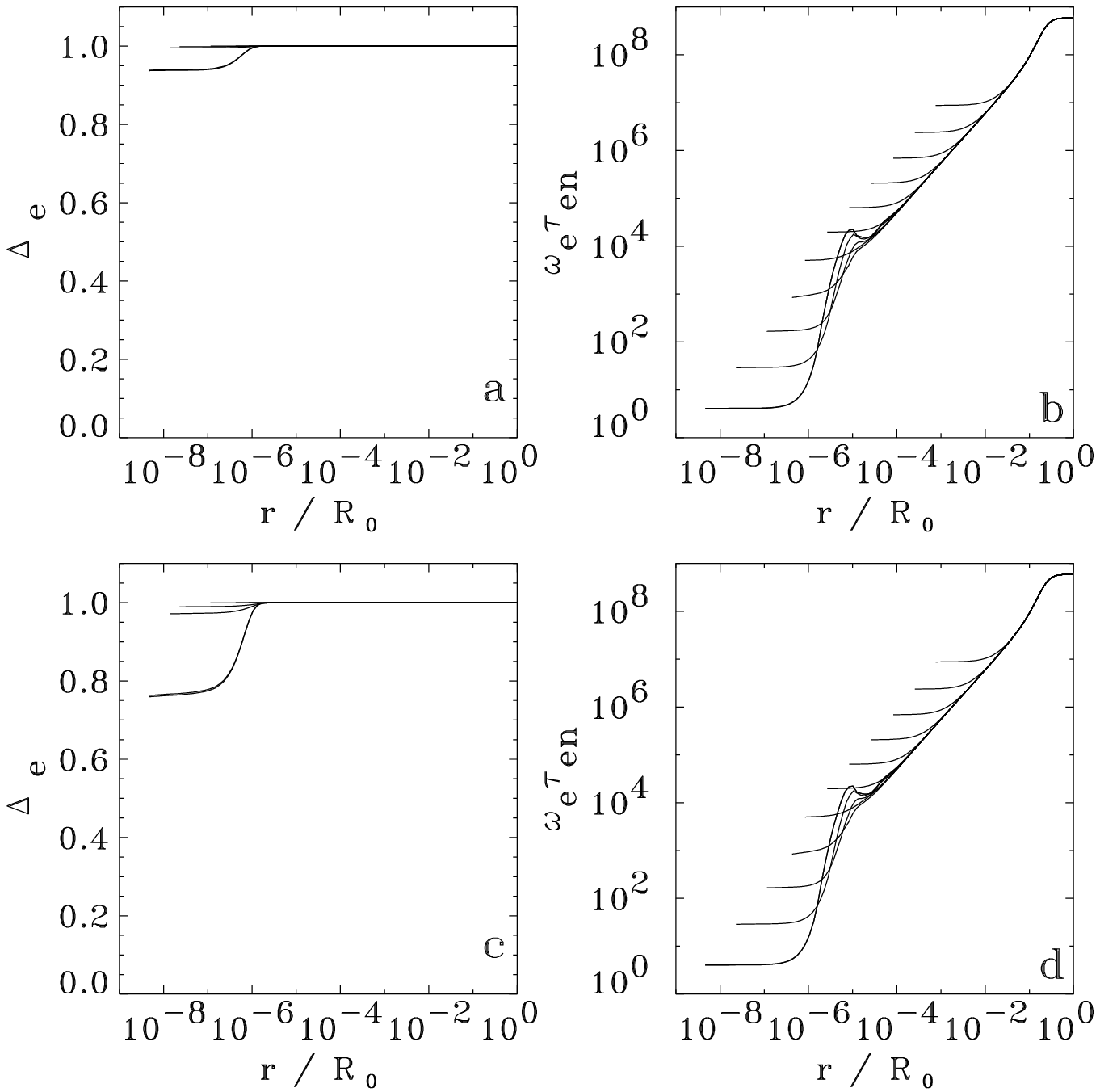}
\caption{\label{al_b} 
Spatial profiles of the indirect (left column) and direct (right column) attachment 
parameters of electrons, in models $m$9 (top row) and $Al$ (bottom row), at different 
times, chosen as in Figure (\ref{nparam_rad_a}).}
\end{figure*}

The greater degree of ionization at high densities affects the degree of attachment of
the electrons on the field lines. Figure (\ref{al_b}) shows radial profiles of 
the indirect ($\Delta_{\rm e}$, left column) and direct
($\omega_{\rm e}\tau_{\rm en}$, right column) attachment parameters of
the electrons, at different values of the central number density of
the neutrals. The top-row plots correspond to model $m$9 and the
bottom-row plots correspond to model $Al$. As we discussed in \S \, \ref{mtfvar},  
the attachment parameter in model $m$9 remains very close to $1$ throughout
the run, and hence the electrons are always well coupled to the
magnetic field. However, in model $Al$ at high
densities, the attachment parameter of the electrons
falls below $0.8$, and hence the electrons have started to detach by the
end of the run. This effect is clearly not of magnetic origin, since the magnetic 
field (and the direct attachment parameter of the electrons at a given density)
is essentially the same in the two models. Instead, the effect is
electrostatic: the increased abundance of positively-charged 
species (positively-charged grains and ions), all of which have completely detached
from the field lines and are following the motion of the neutrals by the densities of
interest, exerts attractive electrostatic forces on the
electrons, which thereby begin to detach earlier than they otherwise would.


\begin{figure*}
\plotone{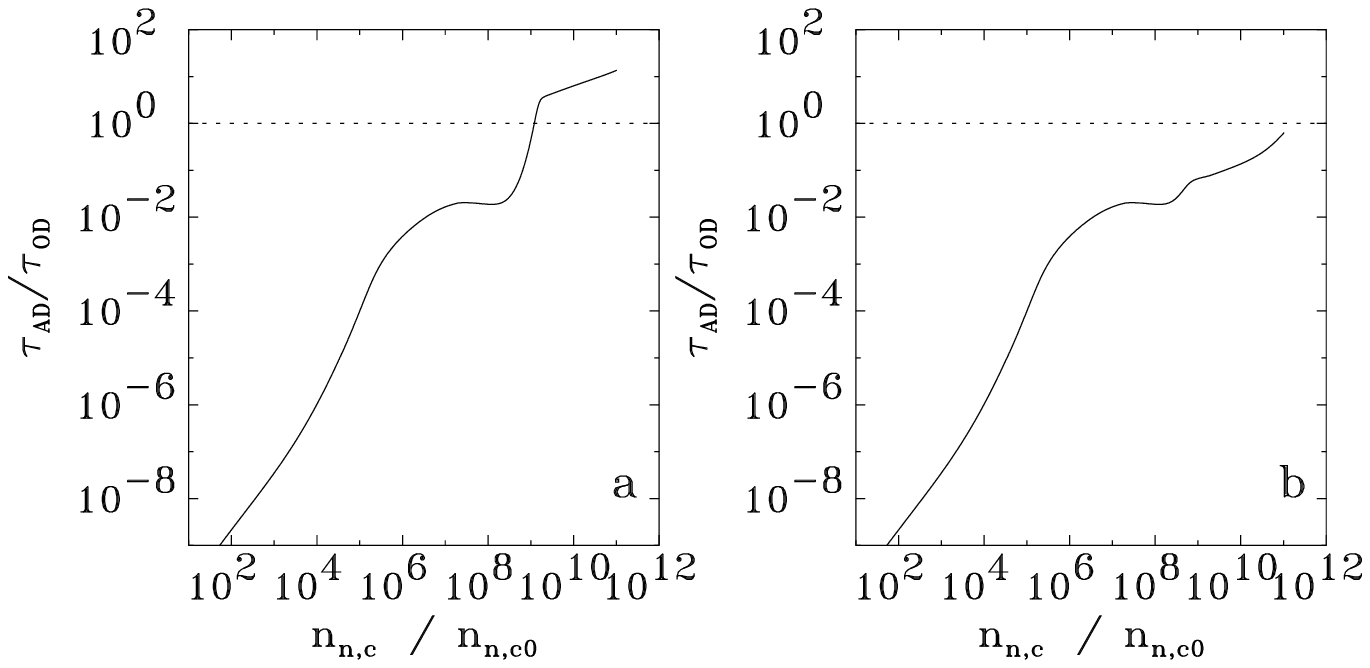}
\caption{\label{al_c} 
Ratio of ambipolar-diffusion and Ohmic-dissipation timescales in
models $m$9 (left) and $Al$ (right) as a function of the central density normalized to 
its initial value ($n_{\rm n,c0} = 4900 \, {\rm cm^{-3}}$).}
\end{figure*}

A second important effect of the higher degree of ionization in model
$Al$ is the difference in the dependence on number density of the ratio
of the ambipolar-diffusion and Ohmic-dissipation timescales, which can be
seen in Figure (\ref{al_c}). The increased degree of
ionization and the better attachment of electrons to field lines allowed ambipolar 
diffusion to operate for higher densities and
hence its timescale to remain smaller than that of Ohmic dissipation
throughout the run.

\section{Conclusions and Discussion}

In Paper II, we presented results for the ambipolar-difussion--initiated formation 
and contraction of a protostellar fragment inside a self-gravitating, magnetically 
supported model molecular cloud, with a fiducial set of values for the free parameters. 
We followed the evolution of the fragment into the opaque regime, through the formation of a 
hydrostatic protostellar core and the later onset of thermal ionization. In this paper, we a
ssessed the sensitivity of the solution on the values of the free parameters; namely, the 
density of neutrals $n_{\rm opq}$ above which the equation of state becomes adiabatic, the 
initial central mass-to-flux ratio $\mu_{\rm d,c0}$ of the parent cloud (relative to its 
critical value for collapse), and the radionuclide that dominates the 
ionization at the highest densities ($n_{\rm n} \gtrsim 10^{12}$ ${\rm cm^{-3}}$).


We found that varying $n_{\rm opq}$ did not result in any qualitatively or 
quantitatively significant change in the structure of the magnetic field inside the 
hydrostatic core. The value of $B_z \approx 0.1 \, {\rm G}$ reached by the end of each 
run (when $T=1000 {\rm \, K}$) is almost the 
same in all cases and, remarkably, very close to the values of the protosolar magnetic 
field as measured in meteorites \citep{levy88,slp61,hr74}. The detachment of the electrons 
from the magnetic field
lines (signifying the decoupling of the magnetic field from the matter) was found to 
occur at $n_{\rm n} \approx 10^{15} \, \, {\rm cm ^{-3}}$, provided that this density 
could be reached in the central part of the hydrostatic core before the temperature reached 
1000 K. If the temperature of 1000 K is reached before the electrons have detached from the 
field lines, thermal ionization will recouple the field lines 
and the matter; hence, complete decoupling will never be achieved. This latter case was 
found to be the most likely outcome, because the electrons did not detach from the field 
lines unless adiabaticity set in only at very high neutral densities ($\gtrsim 10^{12}$ 
${\rm cm^{-3}}$). However, 
a detailed radiative transfer treatment is needed in order to follow the evolution of the 
core temperature more accurately and thereby arrive at a definitive conclusion concerning 
the decoupling of the magnetic field from the matter.

That the structure of the magnetic field, the density at which the electron detachment 
takes place, and the relative importance of ambipolar diffusion and Ohmic dissipation 
in reducing the protostellar magnetic flux are insensitive to the assumed equation of 
state is also corroborated by the control isothermal run (in which the equation of state is 
assumed to remain isothermal throughout the core and for the entire duration of the run). 
However, the onset of adiabaticity and the formation of a central hydrostatic core did facilitate
the development of the magnetic shock, which, in the adiabatic cases, formed at earlier 
times and was significantly stronger. 

The effect of varying the initial mass-to-flux ratio of the parent cloud was found to 
be limited to the time required for the formation of the magnetically supercritical core 
(which is longer for the more subcritical clouds). Beyond that point, the supercritical 
core loses memory of the initial mass-to-flux ratio of the cloud, and the late-time results 
are qualitatively and quantitatively similar in all cases.

Finally, changing the dominant radionuclide (from $^{40}$K to $^{26}$Al) responsible for 
ionization at the late stages of the evolution, ($n_{\rm n} \gtrsim 10^{12}$ ${\rm cm^{-3}}$) 

led to the electrons beginning to detach from the magnetic field lines at slightly smaller 
densities, because the more abundant, detached, infalling positive charges enhanced the 
electrostatic attraction of the electrons. In addition, ambipolar diffusion remained more 
effective than Ohmic dissipation in reducing the magnetic flux of the protostellar core t
hroughout the run.

\acknowledgements{KT would like to thank Vasiliki Pavlidou for useful discussions and Glenn Ciolek 
for providing the base version of the code used in this work.
The work of KT was supported in part by the University of Illinois through a Dissertation 
Completion Fellowship. This research was partially supported by a grant from the American 
Astronomical Society and NSF grants AST 02-06216 and AST 02-39759.}


\begin{thebibliography}

\bibitem[Basu \& Mouschovias (1995)]{BM95a}
Basu, S. \& Mouschovias, T. Ch. 1995, ApJ, 452, 386

\bibitem[Desch \& Mouschovias (2001)]{DM01}
Desch, S. J. \& Mouschovias, T. Ch. 2001, ApJ, 550, 314

\bibitem[Draine \& Lee (1984)]{DL84}
Draine, B. T. \& Lee, H. M. 1984, ApJ, 285, 89

\bibitem[Gaustad (1963)]{Gaustad63}
Gaustad, J. E. 1963, ApJ, 138, 1050

\bibitem[Hayashi (1966)]{Hayas66}
Hayashi, C. 1966, ARA\&A, 4, 171

\bibitem[Herndon \& Rowe (1974)]{hr74}
Herndon, J. M. \& Rowe, M. W. 1974, Meteoritics, 9, 289

\bibitem[Levy (1988)]{levy88}
Levy, E. H. 1988, in Meteorites and the Early Solar System, ed. J. F. Kerridge
  \& M. S. Matthews (Tucson: Univ. Arizona Press)

\bibitem[Masunaga \& Inutsuka (1999)]{MI99}
Masunaga, H. \& Inutsuka, S. 1999, ApJ, 510, 822

\bibitem[Ossenkopf \& Henning(1994)]{OH94}
Ossenkopf, V. \& Henning, T. 1994, A\&A, 291, 943

\bibitem[Stacey, Lovering, \& Parry (1961)]{slp61}
Stacey, F. D., Lovering, J. F., \& Parry, L. G. 1961, J. Geophys. Res., 66,
  1523
\bibitem[Tassis \& Mouschovias (2006)]{TassisM06a}
Tassis, K. \& Mouschovias, T. Ch. 2006a [paper I]

\bibitem[Tassis \& Mouschovias (2006)]{TassisM06b}
---. 2006b [paper II]
\end{thebibliography}
\end{document}